\documentclass{appolb}
\usepackage[english]{babel}
\usepackage{graphicx}
\begin{document}

\title{RELATIVISTIC TWO-PARTICLE EQUATIONS WITH SUPERPOSITION OF DELTA-SHELL
POTENTIALS: SCATTERING AND BOUND STATES
}
\author{Valery Kapshai \thanks{kapshai@tut.by}, Yury Grishechkin \thanks{ygrishechkin@tut.by}
\address{Department of Physics, Gomel State University,
102, Gomel, 246019, Belarus}
}
\maketitle

\markboth{Valery Kapshai, Yury Grishechkin}{Relativistic
two-particle equations}

\begin{abstract}
Exact solutions of two-particle relativistic equations of quantum
field theory describing the scattering $s$-states and the bound
$s$-states are found in the cases of delta-shell potential and
superposition of delta-shell potentials. Some properties of
obtained relativistic wave functions, scattering amplitudes and
quantization conditions are investigated. The resonance character
of scattering processes is demonstrated by the behavior of
amplitudes. It is shown that the non-relativistic limits of these
relativistic values coincide with respective non-relativistic
ones obtained from the Schr\"{o}dinger equation.
\end{abstract}

\section{Introduction}
In this paper we consider three-dimensional relativistic equations
of quasipotential type, which describe systems of two particles with
equal masses. Originally these two-particle equations were
obtained in the momentum representation, in which they are
analogous to the Schr\"{o}dinger and the Lippmann-Schwinger
equations \cite{LogTavkh,KadNP}. In the center of mass system, the
equations for the scattering state wave function of relative movement $\psi_{(j)}(\mathbf{q}, \mathbf{p})$ have the form
\begin{eqnarray}
\label{releq_mr} \psi_{(j)}(\mathbf{q}, \mathbf{p})=(2\pi)^3\delta(\mathbf{p}-\mathbf{q})
+G_{(j)}(E_q,p)\int V(E_q,\mathbf{p},\mathbf{k})\psi_{(j)}(\mathbf{q}, \mathbf{k})\frac{d\mathbf{k}}{(2\pi)^3}.
\end{eqnarray}
In the most general case, the qusipotential $V(E_q,\mathbf{p},\mathbf{k})$
depends on the two-particle energy $2E_q$ in the center of mass
system and on the spin variables. In this paper we only discuss
the case of energy and spin independent potentials. Energy for the
scattering states can be parameterized as follows:
$2E_q=2m\cosh\chi_q$, where $m$ - is the mass of each particle,
and $\chi_q$ - is the rapidity. The Green functions (GFs)
$G_{(j)}(E_q,p)$, in the momentum representation for the
Logunov-Tavkhelidze  equation ($j=1$) and for the Kadyshevsky
equation ($j=2$), have the following form ($E_p=\sqrt{\mathbf{p}^2+m^2}=m\cosh\chi_p$):
\begin{eqnarray}
\label{GF_Log-Tavkh}
G_{(1)}(E_q,p)=\frac{1}{E_q^2-E_p^2+i0}\frac{m}{E_p};\quad
G_{(2)}(E_q,p)=\frac{1}{2E_q-2E_p+i0}\frac{m}{E_p^2}.
\end{eqnarray}
We also consider the modified Logunov-Tavkhelidze ($j=3$) and the
modified Kadyshevsky ($j=4$) equations, for which the GFs are
\begin{eqnarray}
\label{GF_modLog-modTavkh}
G_{(3)}(E_q,p)=\frac{1}{E_q^2-E_p^2+i0};\quad
G_{(4)}(E_q,p)=\frac{1}{2E_q-2E_p+i0}\frac{1}{E_p}.
\end{eqnarray}
All four relativistic Green functions in the non-relativistic
limit ($m\to\infty$; $\chi_q,\chi_p\to 0$; $m\chi_q=q$;
$m\chi_p=p$) turn into the GF of the Schr\"{o}dinger equation:
\begin{equation}
\label{nnrlmGF}\lim\limits_{\stackrel{\chi_q,\chi_p\to 0}{m\to
\infty}} G_{(j)}(E_q,p)=G_{q(0)}(p)=\frac{1}{q^2-p^2+i0}.
\end{equation}

The so called relativistic configurational representation (RCR) is
used for two-particle equations along with the momentum
representation \cite{KadNC,KadSJPN}. The RCR was introduced by
means of expansion of all the values in equations over the
principal series of matrix elements of irreducible unitary
representations the Lorentz group, which are realized by functions
($\mathbf{r}=r\mathbf{n}$)
\begin{equation}
\label{plwav}\xi(\mathbf{p}, \mathbf{r})=\bigg(\frac{E_p-\mathbf{p}\mathbf{n}}{m}\bigg)^{-1-imr}.
\end{equation}
The group parameter $r$ is treated as a relative relativistic
coordinate in the center of mass system \cite{KadNC,KadSJPN}.
Functions $\xi(\mathbf{p}, \mathbf{r})$ play the role of plane waves in the
RCR; in the nonrelativistic limit they turn into functions
$\exp(i\mathbf{p}\mathbf{r})$, where $\mathbf{r}$ - is the radius-vector in the
ordinary coordinate space. For instance, transformation of the
wave function to the RCR and its inverse transformation have the
form \cite{KadNC,KadSJPN}
\begin{equation}
\label{tr_wf} \psi_{(j)}(\mathbf{q},\mathbf{r})=\frac{1}{(2\pi)^3}\int
\xi(\mathbf{p},\mathbf{r})\psi_{(j)}(\mathbf{q},\mathbf{p})\frac{m}{E_p}d\mathbf{p},
\end{equation}
\begin{equation}
\label{invtr_wf} \psi_{(j)}(\mathbf{q},\mathbf{p}\,)=\int\xi^*(\mathbf{p},\mathbf{r})\psi_{(j)}(\mathbf{q},\mathbf{r})d\mathbf{r},
\end{equation}
respectively. Properties of functions $\xi(\mathbf{p}, \mathbf{r})$ and
transformations (\ref{tr_wf}), (\ref{invtr_wf}) as well as
analogous transformations for quasipotentials $V(E_q,\mathbf{p}, \mathbf{k})$ and GFs $G_{(j)}(E_q, p)$ are discussed in detail in\cite{KadNC,KadSJPN}.

After the transformation into the RCR, the equations for
relativistic wave functions have the following form
\cite{KadNC,KadSJPN} (in this paper we discuss only the case of
the RCR-local, spherically symmetric, spin independent potentials
$V(\mathbf{r})\equiv V(r)$):
\begin{eqnarray}
\label{eq_RCR}
\psi_{(j)}(\mathbf{q},\mathbf{r})=\xi(\mathbf{q},\mathbf{r}) +\int
G_{(j)}(E_q,\mathbf{r},\mathbf{r}{\,'})\;V(r{\,'})\;\psi_{(j)}(\mathbf{q},\mathbf{r}{\,'})\;d{\mathbf{r}{\,'}}.
\end{eqnarray}

The expansion of functions $\psi_{(j)}(\mathbf{q}, \mathbf{r})$, $\xi(\mathbf{q},
\mathbf{r})$ into series in the Legendre polynomials
$P_l\big(\mathbf{q}\mathbf{r}/qr\big)$, and of functions  $G_{(j)}(E_q,\mathbf{r}, \mathbf{r'})$ into series in the polynomials
$P_l\big(\mathbf{r}\mathbf{r}'/rr'\big)$, results in equations for the
partial wave functions, which in the $s$-wave case have the form
\begin{eqnarray}
\label{scateq_j} \psi_{(j)}(\chi_q, r)=\sin(\chi_q m r)
+\int\limits_{0}^{\infty} G_{(j)}(\chi_q, r,
r{\,'})\,V(r{\,'})\,\psi _{(j)}(\chi_q, r{\,'})\,{dr{\,'}}.
\end{eqnarray}
Partial Green functions in the RCR $G_{(j)}(\chi_q,r,r{\,'})$ are
connected with the GFs in the momentum representation as follows:
\begin{eqnarray}
\label{fourtrgf} G_{(j)}(\chi_q, r, r{\,'})=
\frac{2}{\pi}\int\limits_{0}^{\infty}\sin(\chi_k mr)
G_{(j)}(m\cosh\chi_q,k)E_k\sin(\chi_k mr{\,'}){d\chi_k}.
\end{eqnarray}

The direct calculation of GFs (\ref{fourtrgf}) for specific $j$
leads to the following expressions:
\begin{equation}
\label{gf_j}G_{(j)}(\chi_q,r,r{\,'})=G_{(j)}(\chi_q,r-r{\,'})-G_{(j)}(\chi_q,r+r{\,'}),
\end{equation}
where \cite{KAJP}
\begin{equation}
\label{onedim_gf} G_{(1)}(\chi_q,r)
=\frac{-i}{K^{(1)}_q}\frac{\sinh[(\pi/2 + i\chi_q)m\;
r]}{\sinh[\pi\;m \;r/2]};
\end{equation}
\begin{displaymath}
G_{(2)}(\chi_q,r)=\frac{(4m\cosh\chi_q)^{-1}}{ \cosh[\pi\; m
\;r/2]}-\frac{i}{K^{(2)}_q}\frac{\sinh[(\pi + i\chi_q)m\;
r]}{\sinh[\pi\; m\; r]};
\end{displaymath}
\begin{displaymath}
G_{(3)}(\chi_q,r)=\frac{-i}{K^{(3)}_q}\frac{\cosh[(\pi/2 + i\chi_q
)m\; r]}{\cosh[\pi\;m \;r/2]};
G_{(4)}(\chi_q,r)=\frac{-i}{K^{(4)}_q}\frac{\sinh[(\pi + i\chi_q
)m\; r]}{\sinh[\pi\;m\; r]}.
\end{displaymath}
In formulae (\ref{onedim_gf}) we used the notations
\begin{equation}
 K^{(1)}_q = K^{(2)}_q = m\;\sinh 2\chi_q;\quad K^{(3)}_q=K^{(4)}_q = 2m\;\sinh \chi_q.
\end{equation}
In what follows, we will need the asymptotics of GFs (\ref{gf_j})
at $r\to\infty$
\begin{equation}
\label{eq_asimpt} \left. G_{(j)}(\chi_q,r,r') \right|_{r
\rightarrow \infty } \cong -\frac{ 2}{K_q^{(j)}}\sin(\chi_q
mr')\exp(i\chi_q mr).
\end{equation}

In the bound state case, equations (\ref{scateq_j}) (as well as
(\ref{eq_RCR})) are modified to the homogeneous form, and the rapidity
$\chi_q$ becomes imaginary ($\chi_q=iw_q$; $0\leq w_q<\pi/2$;
$2E_q=2m\cos w_q$)\cite{KAIzv}:
\begin{equation}
\label{relbseq} \psi_{(j)}(iw_q,r)=\int\limits_{0}^{\infty}
{G_{(j)}(iw_q,r,r\,')\;V({r}\,')\;\psi_{(j)}(iw_q,r\,')\;d{r}\,'}.
\end{equation}

It is not difficult to see that in the nonrelativistic limit all
GFs (\ref{gf_j}) transform into the GF of the three-dimensional
Schr\"{o}dinger equation for $s$-waves in the coordinate
representation \cite{newton, taylor}:
\begin{eqnarray}
\label{nonrgf} \lim\limits_{\stackrel{\chi_q\to 0}{m\to\infty}}
G_{(j)}(\chi_q,r,r'\,)=G_{(0)}(q,r,r\,') =\frac{-1}{q}\sin(q
r_{<})\exp(iq r_{>}).
\end{eqnarray}
The nonrelativistic limit of equations (\ref{scateq_j}),
(\ref{relbseq}) gives the Schr\"{o}dinger equation for scattering
\begin{equation}
\label{nonr_ss} \psi_{(0)}(q,r) = \sin(q\,r)
+\int\limits_{0}^{\infty} G_{(0)}(q,r,r{\,'})\;V(r{\,'})\;\psi
_{(0)}(q,r{\,'})\,d{r{\,'}}
\end{equation}
and bound states
\begin{equation}
\label{nonr_bs} \psi_{(0)}(i\kappa,r) = \int\limits_{0}^{\infty}
G_{(0)}(i\kappa,r,r{\,'})\,V(r{\,'})\,\psi_{(0)}(i\kappa,
r{\,'})\,d{r{\,'}},
\end{equation}
respectively. The quantity $i\kappa$ in equation (\ref{nonr_bs})
is defined as $i\kappa =\lim\limits_{\stackrel{w_q\to
0}{m\to\infty}}imw_q$.

In general, potentials $V$ in the RCR depend on the system energy
$2E_q$. Since the problem of finding relativistic potentials in
the frame of quantum field theory is quite complicated, various
phenomenological potentials are commonly used analogous to the
non-relativistic Schr\"{o}dinger equation. In many cases such
phenomenological potentials are chosen to be independent of
$2E_q$. Moreover, the coordinate dependence of relativistic
potentials is often the same as the coordinate dependence of the
non-relativistic potentials. Even so, the number of cases with
exact solutions to relativistic equations is much smaller than in
the non-relativistic theory. Solutions found for such potentials
and the values obtained based on these solutions allow us to make
definite important conclusions about the general properties of
quasipotential type two-particle equations in the case of more
complicated potentials which do not allow for exact solutions.
The potentials which allow exact solutions and detailed analysis
are, therefore, of special interest. To study the general properties based only on numerical solutions is quite difficult.

Delta-function potentials attract much attention in the
non-relativistic theory
\cite{demkov,albeverio,PhysRevC,JPhysA,APP,PhScr}. As a physical
model, the delta-function potential is used to represent a
potential whose spatial extension is very small compared to all
other length scales. Superposition and array of delta-function
potentials have been used in solid state physics and optics. The
solution of one-dimensional relativistic two-particle equations
with the delta-function potential and superposition of
delta-function potentials was considered in \cite{KAJP, KAIzv}.
These equations with potential "delta-function derivative of
$n$-th order" $\delta^{(n)}$ at $n=1,2,3$ were solved in article
\cite{KGRPJ}. The relativistic one-dimensional problem for
nonlinear delta-function potential was considered in
\cite{KHNPCS}.

In this paper we consider the solution of relativistic partial
two-particle equations for $s$-waves (\ref{scateq_j}),
(\ref{relbseq}) with the potential $V_0\delta(r-a)$, which is
localized on the sphere of finite radius $a>0$ (in the
three-dimensional case such potential is often called
\textquotedblleft delta-shell potential\textquotedblright), and
with a superposition of two such potentials. The article is
organized as follows. In the second paragraph the wave functions
for the scattering states are found. The scattering amplitudes,
$S$-matrices and phase shifts are obtained using the wave
functions. The analysis of obtained results is carried out. In
the third paragraph the wave functions and the energy
quantization conditions for them are found for bound states. In
the fourth paragraph the results are studied in the
non-relativistic limit.

\section{The scattering states}

Let us consider briefly the solutions of relativistic equations
for the scattering states (\ref{scateq_j}) in the case of the
delta-shell potential
\begin{equation}
\label{dlt_ptn} V(r)=V_0\delta(r-a),
\end{equation}
where $V_0$ and $a>0$ - are real constants. The wave functions for
potential (\ref{dlt_ptn}) are
\begin{equation}
\label{wf1_j} \psi_{(j)}(\chi_q,r)=\sin(\chi_q m r)
+V_0A_{(j)}^{-1}(\chi_q)\sin(\chi_q m a)G_{(j)}(\chi_q,r,a);
\end{equation}
\begin{displaymath}
 A_{(j)}(\chi_q)=1-V_0G_{(j)}(\chi_q,a,a).
\end{displaymath}

Taking into account (\ref{eq_asimpt}) one can represent the
asymptotics of wave functions (\ref{wf1_j}) at $r\to\infty$ in the
form ($q=m\sinh\chi_q$ - is the relativistic momentum)
\begin{equation}
\label{wf1_asimpt} \left. \psi_{(j)}(\chi_q,r) \right|_{r
\rightarrow \infty } \cong\sin(\chi_q m r)+q
f_{(j)}(\chi_q)\exp(i\chi_q m r),
\end{equation}
\begin{equation}
\label{ampl1_j}f_{(j)}(\chi_q)=\frac{-2V_0\sin^2(\chi_q m a)}{q
K^{(j)}_q A_{(j)}(\chi_q)},
\end{equation}
where the relativistic scattering amplitude $f_{(j)}(\chi_q)$ is
determined, in the same way as the non-relativistic one, as the
coefficient at the scattered wave divided by momentum
\cite{newton, taylor} (the relativistic scattered $s$-wave has the
form $\exp(i\chi_q m r)$ \cite{KadNC, KadSJPN}). Expressions for
the scattering amplitudes corresponding to the four GF variants
can be represented as
\begin{equation}
\label{ampl}f_{(1)}(\chi_q)=\frac{-2V_0q^{-1}s^2(a)}{K^{(1)}_q+V_0\displaystyle\mathstrut\Bigg[\frac{s(2a)}{\tilde{t}(a)}-\frac{2\chi_q}{\pi}+2i
s^2(a)\Bigg]};
\end{equation}
\begin{displaymath}
f_{(2)}(\chi_q)=\frac{-2V_0q^{-1}s^2(a)}{K^{(2)}_q+V_0\displaystyle\mathstrut\Bigg[\frac{
s(2a)}{\tilde{t}(2a)}-\frac{\chi_q}{\pi}+2is^2(a)-\frac{\tilde{t}^2(a/2)}{1+\tilde{t}^2(a/2)}\sinh(\chi_q)\Bigg]};
\end{displaymath}
\begin{displaymath}
f_{(3)}(\chi_q)=\frac{-2V_0q^{-1}s^2(a)}{K^{(3)}_q+V_0\big[\tilde{t}(a)s(2a)+2i
s^2(a)\big]};
\end{displaymath}
\begin{displaymath}
f_{(4)}(\chi_q)=\frac{-2V_0q^{-1}s^2(a)}{K^{(4)}_q+V_0\displaystyle\mathstrut\Bigg[\frac{s(2a)}{\tilde{t}(2a)}-\frac{\chi_q}{\pi}+2i
s^2(a)\Bigg]}.\nonumber
\end{displaymath}
where we used notations
\begin{eqnarray}
\label{ntts2}s(a)=\sin(\chi_qma),\quad \tilde{t}(a)=\tanh(\pi ma).
\end{eqnarray}
Some numerical calculations results of scattering amplitudes
(\ref{ampl1_j}) for $j=1,3$ will be shown later, namely after
considering another important value --- the phase shift.

Now let us find the solutions of equations (\ref{scateq_j}) and
the expressions for the scattering amplitudes in the case of
superposition of two delta-shell potentials:
\begin{equation}
\label{dlt_ptn2}V(r)=V_1\delta(r-a_1)+V_2\delta(r-a_2),
\end{equation}
where $V_{1,2}$, $a_{1,2}$ - are real constants and $a_2>a_1>0$.
Substitution of (\ref{dlt_ptn2}) into equations (\ref{scateq_j})
gives the wave function expressions in the following general form:
\begin{equation}
\label{gnfm_wf}\psi_{(j)}(\chi_q,r)=\sin(\chi_q m
r)+\sum_{k=1}^{2} V_kG_{(j)}(\chi_q,r,a_k)\psi_{(j)}(\chi_q,a_k),
\end{equation}
where the values $\psi_{(j)}(\chi_q,a_{1,2})$ are to be found. To
determine these values one should take expressions (\ref{gnfm_wf})
at the points $r=a_1$ and $r=a_2$. As a result, one obtains a
system of two linear algebraic equations for
$\psi_{(j)}(\chi_q,a_1)$, $\psi_{(j)}(\chi_q,a_2)$. Solving this
system and substituting the solutions into (\ref{gnfm_wf}) one
obtains the following expressions for the wave functions:
\begin{equation}
\label{wf2_j}\psi_{(j)}(\chi_q,r)=\sin(\chi_q m
r)+\sum_{k=1}^{2}V_kG_{(j)}(\chi_q,r,a_k)\frac{\Delta_{(j)k}(\chi_q)}{\Delta_{(j)}(\chi_q)},
\end{equation}
where notations
\begin{equation}
\label{nottns}\Delta_{(j)}(\chi_q)=\prod_{k=1}^{2}[1-V_kG_{(j)}(\chi_q,a_k,a_k)]-V_1V_2G_{(j)}^2(\chi_q,a_1,a_2);
\end{equation}
\begin{displaymath}
\Delta_{(j)1}(\chi_q)=s(a_1)[1-V_2G_{(j)}(\chi_q,a_2,a_2)]+V_2s(a_2)G_{(j)}(\chi_q,a_1,a_2);
\end{displaymath}
\begin{displaymath}
\Delta_{(j)2}(\chi_q)=s(a_2)[1-V_1G_{(j)}(\chi_q,a_1,a_1)]+V_1s(a_1)G_{(j)}(\chi_q,a_1,a_2)
\end{displaymath}
are introduced. Asymptotic behaviour at $r\to\infty$ of wave
functions (\ref{wf2_j}) yields the following formulae for the
scattering amplitudes:
\begin{equation}
\label{ampl2_j}f_{(j)}(\chi_q)=\frac{-2}{q K^{(j)}_q
\Delta_{(j)}(\chi_q)}\sum_{k=1}^{2}V_k\Delta_{(j)k}(\chi_q)s(a_k).
\end{equation}

The explicit form of expressions (\ref{wf2_j})-(\ref{ampl2_j}) for
concrete $j$ in the case of superposition of delta-shell
potentials is quite cumbersome. For example, the scattering
amplitude at $j=3$ is
\begin{equation}
\label{ampl12_3}f_{(3)}(\chi_q)=\frac{F_1+F_2}{qK^{(3)}_q(F_3+iF_4)},
\end{equation}
where
\begin{equation}
\label{F_1234} F_1=V_1s^2(a_1)\Bigg[1+\frac{V_2}{K^{(3)}_q}\tilde{t}(
a_2)s(2a_2)\Bigg]+V_2s^2(a_2)\Bigg[1+\frac{V_1}{K^{(3)}_q}\tilde{t}(a_1)s(2a_1)\Bigg];
\end{equation}
\begin{displaymath}
F_2=\frac{2V_1V_2}{K^{(3)}_q}s(a_1)s(a_2)\Bigg[\tilde{t}\bigg(\frac{a_2-a_1}{2}\bigg)s(a_2-a_1)-\tilde{t}\bigg(\frac{a_2+a_1}{2}\bigg)s(a_2+a_1)\Bigg];
\end{displaymath}
\begin{displaymath}
F_3=\prod_{k=1}^{2}\Bigg[1+\frac{V_k}{K^{(3)}_q}\tilde{t}(a_k)s(2a_k)\Bigg]
\end{displaymath}
\begin{displaymath}
-\frac{V_1V_2}{\big(K^{(3)}_q\big)^2}\Bigg[\tilde{t}\bigg(\frac{a_2-a_1}{2}\bigg)s(a_2-a_1)-\tilde{t}\bigg(\frac{a_2+a_1}{2}\bigg)s(a_2+a_1)\Bigg]^2;
\end{displaymath}
\begin{displaymath}
 F_4=\frac{2V_1}{K^{(3)}_q}s^2(a_1)\Bigg[1+\frac{V_2}{K^{(3)}_q}\tilde{t}(a_2)s(2a_2)\Bigg]+\frac{2V_2}{K^{(3)}_q}s^2(a_2)\Bigg[1+\frac{V_1}{K^{(3)}_q}\tilde{t}(a_1)s(2a_1)\Bigg]
\end{displaymath}
\begin{displaymath}
+\frac{4V_1V_2}{\big(K^{(3)}_q\big)^2}s(a_1)s(a_2)\Bigg[\tilde{t}\bigg(\frac{a_2-a_1}{2}\bigg)s(a_2-a_1)-\tilde{t}\bigg(\frac{a_2+a_1}{2}\bigg)s(a_2+a_1)\Bigg].
\end{displaymath}
Results of numerical calculations of scattering amplitudes
(\ref{ampl2_j}) for $j=1,2,3,4$ will be presented after the phase
shift consideration.

The scattering amplitude $f_{(j)}(\chi_q)$ provides information
about particle scattering. For instance, the partial $s$-wave
cross section $\sigma_{0(j)}(\chi_q)$ and the partial-wave
$S$-matrix $S_{(j)}(\chi_q)$ are expressed through the scattering
amplitude $f_{(j)}(\chi_q)$ by relations
\begin{equation}
\label{cr_sect}\sigma_{0(j)}(\chi_q)=4\pi|f_{(j)}(\chi_q)|^2;\quad
S_{(j)}(\chi_q)=1+2iqf_{(j)}(\chi_q).
\end{equation}

It is not difficult to see that expressions for scattering
amplitudes (\ref{ampl1_j}), (\ref{ampl2_j}) satisfy the unitarity
condition \cite{KadNC, KadSJPN}, which has the form similar to the
non-relativistic one \cite{newton, taylor}:
\begin{equation}
\label{unit_cond} \textup{Im}f_{(j)}(\chi_q)=q|f_{(j)}(\chi_q)|^2.
\end{equation}
It follows from expressions (\ref{ampl1_j}), (\ref{ampl2_j}) that
this unitarity condition is equivalent to the following property
of GFs (\ref{gf_j}):
\begin{equation}
\label{unitconf_GF}\textup{Im}G_{(j)}(\chi_q,a,b)=-\frac{2s(a)s(b)}{K^{(j)}_q}.
\end{equation}

In the case of delta-shell potential (\ref{dlt_ptn}) the partial
wave $S$-matrix can be represented in the form
\begin{equation}
\label{Smatr1}S_{(j)}(\chi_q)=1-\frac{4iV_0s^2(a)}{K^{(j)}_qA_{(j)}(\chi_q)}=\frac{\big(A_{(j)}(\chi_q)\big)^*}{A_{(j)}(\chi_q)},
\end{equation}
from which its unitarity is obvious. For the superposition of
delta-shell potentials (\ref{dlt_ptn2}) $S_{(j)}(\chi_q)$ can also
be represented in analogous form to (\ref{Smatr1})
\begin{equation}
\label{Smatr2}S_{(j)}(\chi_q)=1-\frac{4i\bigg(V_1s(a_1)\Delta_{(j)1}+V_2s(a_2)\Delta_{(j)2}\bigg)}{K^{(j)}_q\Delta_{(j)}}
=\frac{\big(\Delta_{(j)}\big)^*}{\Delta_{(j)}}.
\end{equation}
The unitarity of the $S$-matrix is reflected in the following
representation
\begin{equation}
\label{Smatrunit}S_{(j)}(\chi_q)=\exp(2i\phi_{(j)}(\chi_q)),
\end{equation}
which defines the phase shift $\phi_{(j)}(\chi_q)$. The phase
shifts can be found from expressions:
\begin{enumerate}
\item[a)] in the case of delta-shell potential
\begin{equation}
\label{phsh1}\tan(2\phi_{(j)})=\frac{-4V_0K^{(j)}_q[1-V_0\textup{Re}G_{(j)}(\chi_q,a,a)]s^2(a)}{[K^{(j)}_q(1-V_0\textup{Re}G_{(j)}(\chi_q,a,a))]^2-[2V_0s^2(a)]^2};
\end{equation}
\item[b)] in the case of delta-shell
potential superposition
\begin{equation}
\label{phsh2}\tan(2\phi_{(j)})=\frac{-2\textup{Re}\Delta_{(j)}(\chi_q)\textup{Im}\Delta_{(j)}(\chi_q)}{(\textup{Re}\Delta_{(j)}(\chi_q))^2-(\textup{Im}\Delta_{(j)}(\chi_q))^2},
\end{equation}
\end{enumerate}
where
\begin{equation} \label{ReImDelta}\textup{Re}\Delta_{(j)}(\chi_q)=\prod_{k=1}^{2}[1-V_k\textup{Re}G_{(j)}(\chi_q,a_k,a_k)]
-V_1V_2\Big[\textup{Re}G_{(j)}(\chi_q,a_1,a_2)\Big]^2;
\end{equation}
\begin{displaymath}
\textup{Im}\Delta_{(j)}(\chi_q)=\frac{2V_1s^2(a_1)}{K^{(j)}_q}[1-V_2\textup{Re}G_{(j)}(\chi_q,a_2,a_2)]
\end{displaymath}
\begin{displaymath}
+\frac{2V_2s^2(a_2)}{K^{(j)}_q}[1-V_1\textup{Re}G_{(j)}(\chi_q,a_1,a_1)]
+\frac{4V_1V_2s(a_2)s(a_1)}{K^{(j)}_q}\textup{Re}G_{(j)}(\chi_q,a_1,a_2).
\end{displaymath}

\begin{figure}
    \centering
        \includegraphics[scale=0.7]{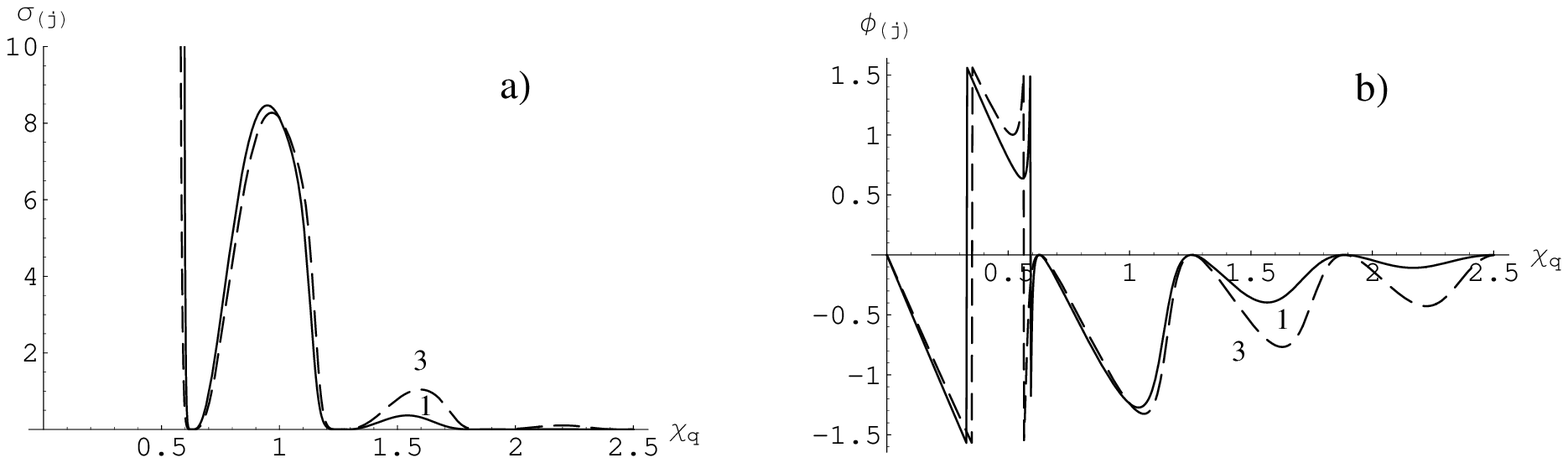}
    \caption{The cross sections (a) and the phase shifts (b) for the delta-shell potential at $m=1$, $a=5$, $V_0=2$}
    \label{fig:fig1}
\end{figure}

\begin{figure}
    \centering
        \includegraphics[scale=0.7]{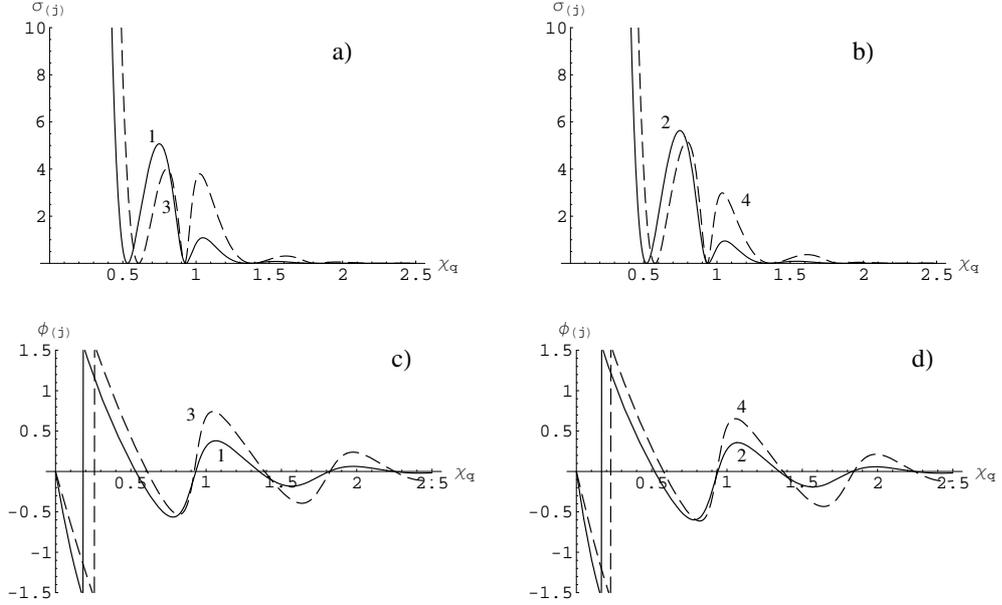}
    \caption{The cross sections (a, b) and the phase shifts (c, d) for the superposition of delta-shell potentials at $m=1$, $a_1=3$, $a_2=4$, $V_1=1$, $V_2=-1$}
    \label{fig:fig2}
\end{figure}

\begin{figure}
    \centering
        \includegraphics[scale=0.7]{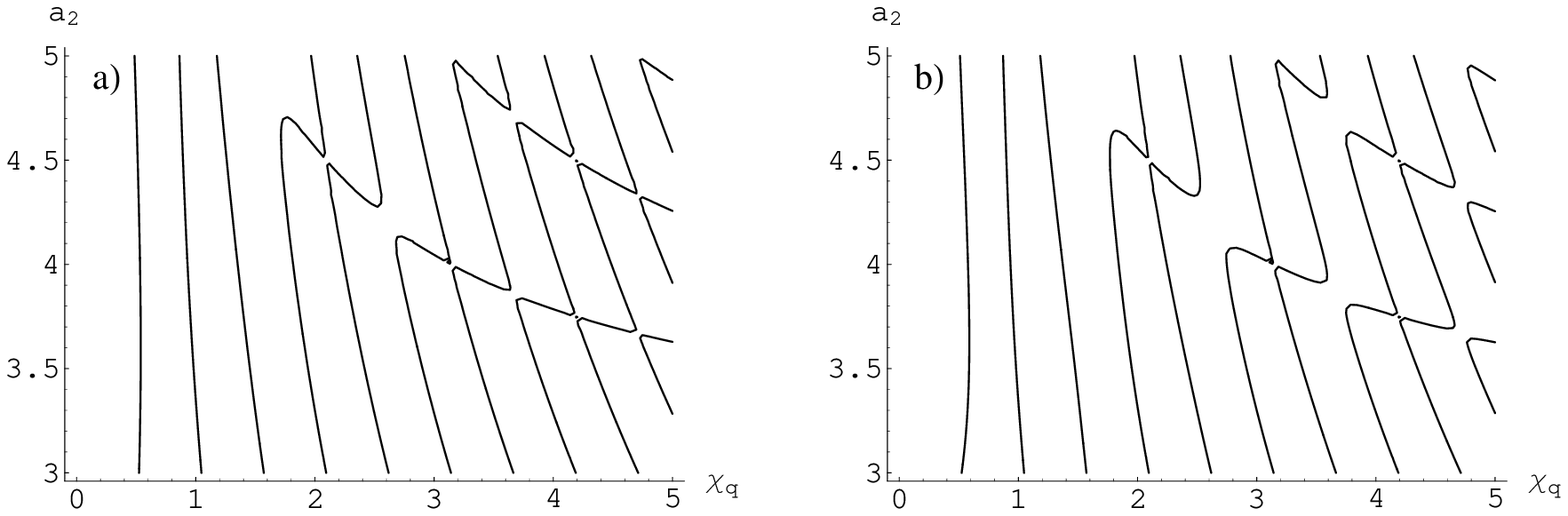}
    \caption{The scattering amplitudes vanishing conditions at $m=1$, $a_1=3$, $V_1=1$, $V_2=-1$: a) $j=1$, b) $j=4$}
    \label{fig:fig3}
\end{figure}

In figures \ref{fig:fig1} and \ref{fig:fig2} we present the
results of numerical calculations of the cross sections and phase
shifts (the curve number corresponds to the index of GF $j$).
In the figures (and from expressions (\ref{ampl1_j}),
(\ref{ampl2_j})) it is seen that $\sigma_{0(j)}(\chi_q)\to 0$ at
$\chi_q\to\infty$, which is natural. The figures also show that
the amplitudes can be equal to zero at some finite values of
rapidity $\chi_q$. The analogous non-relativistic effect of the
partial-wave cross section (scattering amplitude) vanishing at
finite values of momentum is called the Ramzauer-Taunsend effect
\cite{newton,taylor,ShTr}. Scattering amplitudes of the
delta-shell potential are equal to zero if the following condition
(the same for all $j$) holds:
\begin{equation}
\label{RTeff1_j}\chi_qma=\pi n,\quad n=1,2,3,...\,.
\end{equation}
From this condition it follows that the scattering amplitudes have
an infinite number of zeros. For the superposition of two
delta-shell potentials the vanishing condition of the cross
section is
\begin{equation}
\label{RTeff2_j}V_2s^2(a_2)+V_1s^2(a_1)+V_1V_2\big[2s(a_2)s(a_1)G_{(j)}(\chi_q,a_1,a_2)
\end{equation}
\begin{displaymath}
-s^2(a_1)G_{(j)}(\chi_q,a_2,a_2)-s^2(a_2)G_{(j)}(\chi_q,a_1,a_1)\big]=0.
\end{displaymath}
For example, for $j=3$ expression (\ref{RTeff2_j}) has the form
($c(a)=\cos(\chi_q m a)$)
\begin{equation}
\label{RTeff2_3}V_2s^2(a_2)+V_1s^2(a_1)+\frac{2V_1V_2}{K^{(3)}_q}s(a_2)s(a_1)\bigg[\tilde{t}\bigg(\frac{a_2-a_1}{2}\bigg)s(a_2-a_1)
\end{equation}
\begin{displaymath}
-\tilde{t}\bigg(\frac{a_2+a_1}{2}\bigg)s(a_2+a_1)+\tilde{t}(a_1)s(a_2)c(a_1)+\tilde{t}(a_2)s(a_1)c(a_2)\bigg]=0.
\end{displaymath}
Condition (\ref{RTeff2_j}) depends on five variables ($V_1$,
$V_2$, $a_1$, $a_2$, $\chi_q$). When three of the variables are
fixed, a condition for the other two is obtained. In this
2-variable case the scattering amplitude zeros can be represented
on a plane. Some results of numerical calculations (for
expressions (\ref{RTeff2_j})) of this sort are represented in
figure \ref{fig:fig3}. At $\chi_q\to\infty$ expression
(\ref{RTeff2_j}) turns into equality
\begin{equation}
\label{condinf}V_2\sin^2(\chi_qma_2)+V_1\sin^2(\chi_qma_1)=0.
\end{equation}
From (\ref{condinf}) it follows:
\begin{enumerate}
\item[1)] at $V_1V_2<0$ there is an infinite number of zeros of the
scattering amplitude $f_{(j)}(\chi_q)$; \item[2)] at $V_1V_2>0$
condition (\ref{condinf}) only has solutions if the quotient
$a_2/a_1$ is a rational number, in such case there is an infinite
number of scattering amplitude zeros, otherwise the scattering
amplitude $f_{(j)}(\chi_q)$ has a finite number of zeros.
\end{enumerate}

\section{The bound states}

Let us consider solutions of equations for the bound states
(\ref{relbseq}) (for all $j$) with potentials (\ref{dlt_ptn}) and
(\ref{dlt_ptn2}). The wave functions of bound states for
delta-shell potential (\ref{dlt_ptn}) have the form
\begin{equation}
\label{wfbs1_j}
\psi_{(j)}(iw_q,r)=V_0G_{(j)}(iw_q,r,a)\psi_{(j)}(iw_q,a).
\end{equation}
Nontrivial $\psi_{(j)}(iw_q,a)$ (not equal to zero) exist only if
the following conditions hold:
\begin{equation}
\label{qncn1_j}V_{0(j)}=[G_{(j)}(iw_q,a,a)]^{-1}=[G_{(j)}(iw_q,0)-G_{(j)}(iw_q,2a)]^{-1},
\end{equation}
these are the quantization conditions for the quantity $w_q$
($2E_q=2m\cos w_q$). In conditions (\ref{qncn1_j}) the potential
parameter $V_0$ is represented as a function of the energy
parameter $w_q$. Of course, it would be more convenient to have
$w_q$ (or $E_q$) as a function of $V_0$, but this is impossible in the case considered here. For $j=1,2,3,4$ conditions (\ref{qncn1_j}) have the following form:
\begin{equation}
\label{qncn1_1}V_{0(1)}=\frac{\pi
m\sin2w_q}{2w_q+\pi\displaystyle\mathstrut\Bigg[2\tilde{s}^{\,2}(a)-\frac{\tilde{s}(2a)}{\tilde{t}(a)}\Bigg]};
\end{equation}
\begin{displaymath}
V_{0(2)}=\frac{\pi
m\sin2w_q}{w_q+\pi\displaystyle\mathstrut\Bigg[\frac{\tilde{t}^2(a/2)\sin
w_q}{\tilde{t}^2(a/2)+1}+2\tilde{s}^{\,2}(a)-\frac{\tilde{s}(2a)}{\tilde{t}(2a)}\Bigg]};
\end{displaymath}
\begin{displaymath}
V_{0(3)}=\frac{2m\sin
w_q}{2\tilde{s}^{\,2}(a)-\tilde{t}(a)\tilde{s}(2a)};\quad
V_{0(4)}=\frac{2\pi m\sin
w_q}{w_q+\pi\displaystyle\mathstrut\Bigg[2\tilde{s}^{\,2}(a)-\frac{\tilde{s}(2a)}{\tilde{t}(2a)}\Bigg]},
\end{displaymath}
where we used notations
\begin{equation}
\label{nts3} \tilde{s}(a)=\sinh(w_q m a);\quad
\tilde{t}(a)=\tanh(\pi ma).
\end{equation}
\begin{figure}[b!]
    \centering
        \includegraphics[scale=0.7]{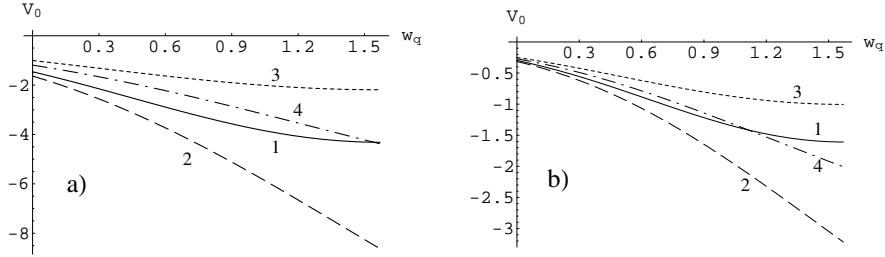}
    \caption{The quantization condition for the delta-shell potential at: a) $m=1$, $a=1$; b) $m=0.5$, $a=2$}
    \label{fig:fig4}
\end{figure}
From expressions (\ref{qncn1_1}) it follows that bound states can
exist only if $V_{0(j)}<0$. In figure \ref{fig:fig4} we present
numerical calculation results of $V_0$, $w_q$ -dependence
(for all $j$) given by (\ref{qncn1_1}). As seen in the figure,
only one energy level can exist for the delta-shell potential (one
value $w_q$ with all the other parameters fixed). Unknown values
$\psi_{(j)}(iw_q,a)$ in wave functions (\ref{wfbs1_j}) can be
determined from the normalization condition which has the same
form in the RCR for all $j$:
\begin{equation}
\label{norm_cond}\int_0^\infty|\psi_{(j)}(iw_q,r)|^2{dr}=1.
\end{equation}
Solutions of equations (\ref{relbseq}) for the superposition
of the delta-shell potentials can be found by analogy with the
scattering state case. The quantization conditions for the
superposition of delta-shell potentials (\ref{dlt_ptn2}) are
\begin{equation}
\label{qncn2_j}
\prod_{k=1}^{2}[1-V_kG_{(j)}(iw_q,a_k,a_k)]-V_1V_2G_{(j)}^2(iw_q,a_1,a_2)=0.
\end{equation}

\begin{figure}
    \centering
        \includegraphics[scale=0.7]{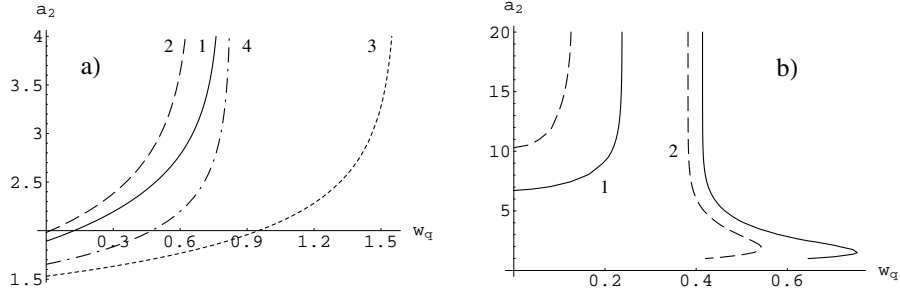}
    \caption{The quantization condition for the superposition of two delta-shell potentials at $m=1$, $a_1=1$,: a) $V_1=7$, $V_2=-2$; b) $V_1=-2$, $V_2=-1$}
    \label{fig:fig5}
\end{figure}

\begin{figure}[h!]
    \centering
        \includegraphics[scale=0.7]{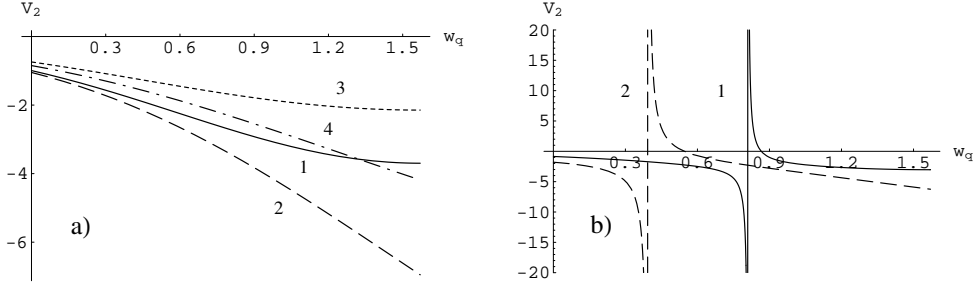}
    \caption{The quantization condition for the superposition of two delta-shell potentials at $m=1$, $a_1=1$: a) $a_2=2$, $V_1=1.5$; b) $a_2=4$, $V_1=-3.5$}
    \label{fig:fig6}
\end{figure}

It is not difficult to see that  at $a_1\to\infty$ (or
$a_2\to\infty$) expressions (\ref{qncn2_j}) are transformed into
quantization conditions similar to (\ref{qncn1_j}) for the single
delta-shell. In figure \ref{fig:fig5} we present some numerical
results for (\ref{qncn2_j}). The figure shows that one or two
energy levels can exist for the superposition of two delta-shell
potentials (one or two values $w_q$ for fixed $a_2$).
Expressing $V_2$ via all other parameters of the problem under
consideration ($V_1$, $a_1$, $a_2$, $m$, $w_q$) one obtains
\begin{equation}
\label{V2}V_{2(j)}=\frac{1-V_1G_{(j)}(iw_q,a_1,a_1)}{G_{(j)}(iw_q,a_2,a_2)+V_1F_{(j)}(iw_q,a_1,a_2)},
\end{equation}
\begin{displaymath}
F_{(j)}(iw_q,a_1,a_2)=G_{(j)}^2(iw_q,a_1,a_2)-\prod_{s=1}^{2}G_{(j)}(iw_q,a_s,a_s).
\end{displaymath}
In figure \ref{fig:fig6} we show some numerical results of
expressions (\ref{V2}) at fixed parameters $V_1$, $a_1$, $a_2$,
$m$. The figure illustrates that expressions (\ref{V2}) can have
singularities at some parameters $V_1$, $a_1$, $a_2$, $m$, $w_q$.

\begin{figure}%[t!]
    \centering
        \includegraphics[scale=0.8]{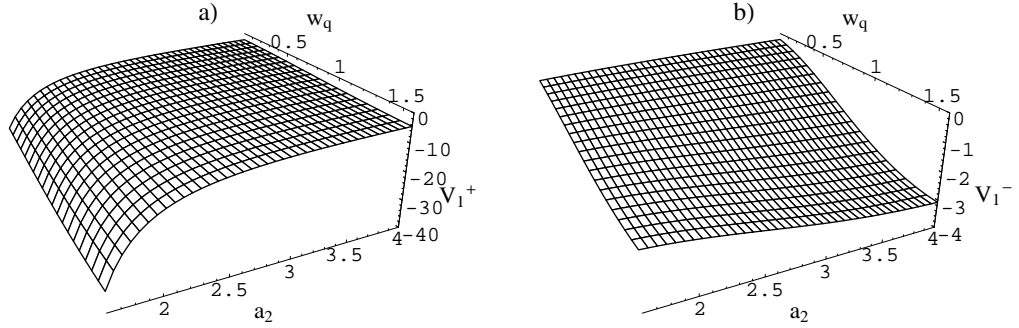}
    \caption{The quantization condition for the superposition of delta-shell potentials for $j=1$ at $\alpha=1$, $m=1$, $a_1=1$}
    \label{fig:fig7}
\end{figure}

\begin{figure}
    \centering
        \includegraphics[scale=0.8]{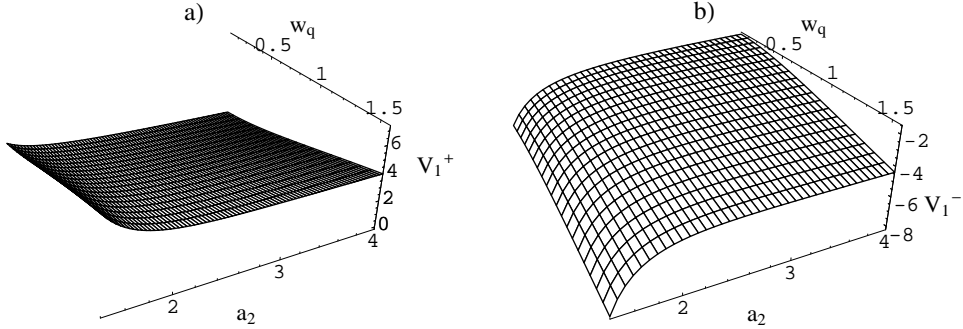}
    \caption{The quantization condition for the superposition of delta-shell potentials for $j=4$ at $\alpha=-1$, $m=1$, $a_1=1$}
    \label{fig:fig8}
\end{figure}

For a more detailed analysis of expressions (\ref{qncn2_j}) let us
define $V_2=\alpha V_1$, where $\alpha$ is a dimensionless
parameter ($\alpha\neq 0$). Equalities (\ref{qncn2_j}) then take
the form of quadratic equations for $V_{1}$, solving them one can
obtain two following expressions for each $j$:
\begin{equation}
\label{qncn2_sc} V_{1(j)}^{\pm}=\frac{\alpha
G_{(j)}(iw_q,a_2,a_2)+G_{(j)}(iw_q,a_1,a_1)\pm
\sqrt{D_{(j)}}}{2\alpha
\big[G_{(j)}(iw_q,a_1,a_1)G_{(j)}(iw_q,a_2,a_2)-G_{(j)}^2(iw_q,a_1,a_2)\big]},
\end{equation}
\begin{displaymath}
D_{(j)}=\big[G_{(j)}(iw_q,a_1,a_1)-\alpha
G_{(j)}(iw_q,a_2,a_2)\big]^2+4\alpha G_{(j)}^2(iw_q,a_1,a_2).
\end{displaymath}
Numerical results for expressions (\ref{qncn2_sc}) at $V_1=V_2$
($\alpha=1$) and at $V_1=-V_2$ ($\alpha=-1$) are shown in figures
\ref{fig:fig7} and \ref{fig:fig8} respectively. From the
figures and expressions (\ref{qncn2_sc}) it follows that two values $V_1$ exist
for each value $w_q$ at fixed parameters $m$, $a_1$, $a_2$,
$\alpha$. Herewith, bound states can exist if one of the following conditions
hold: 1) $V_1<0$, $V_2<0$; 2) $V_1<0$, $V_2>0$; 3) $V_1>0$, $V_2<0$.
At $V_1>0$, $V_2>0$ bound states do not exist.

It has to be noted that quantization conditions (\ref{qncn1_j})
and (\ref{qncn2_j}) can be obtained by equating the expressions for
$A_{(j)}(\chi_q)$, $\Delta_{(j)}(\chi_q)$ (expressions (\ref{wf1_j}) and
(\ref{nottns})) to zero at $\chi_q=iw_q$, $0\leq w_q<\pi/2$.

\section{The non-relativistic limit}
Now let us consider the non-relativistic limit of the results
obtained. In this limit the wave functions, the scattering
amplitudes and the quantization conditions obtained give the same
result for all equations under consideration. In the case of
single delta-potential the limits are ($q$ - is the non-relativistic momentum, and $\kappa =\lim\limits_{\stackrel{w_q\to0}{m\to\infty}}mw_q$.)
\begin{eqnarray}
\label{nonr_wf} \lim\limits_{\stackrel{\chi_q\to
0}{m\to\infty}}\psi_{(j)}(\chi_q,r)=\psi_{(0)}(q,r)
=\sin(qr)+\frac{V_0q\sin(qa)G_{(0)}(q,r,a)}{q+V_0\sin(qa)\exp(iqa)},
\end{eqnarray}
\begin{equation}
\label{nonr_ampl} \lim\limits_{\stackrel{\chi_q\to
0}{m\to\infty}}f_{(j)}(\chi_q)=f_{(0)}(q)=\frac{-V_0\sin^2(qa)}{q[q+V_0\sin(qa)\exp(iqa)]},
\end{equation}
\begin{equation}
\label{nonr_qncn} \lim\limits_{\stackrel{w_q\to
0}{m\to\infty}}V_{0(j)}(w_q)=V_{0(0)}(\kappa)=\frac{-2\kappa}{1-\exp(-2\kappa
a)}.
\end{equation}
Expressions (\ref{nonr_wf})-(\ref{nonr_qncn}) coincide with the
corresponding results obtained based on the Sch\"{o}dinger
equation solution \cite{landau,gottfried,bolsterli}. In the case
of superposition of two delta-shell potentials the
non-relativistic limits of the relativistic wave functions and
scattering amplitudes for all $j$ can be written as
\begin{equation}
\label{nonr_wf2}
\psi_{(0)}(q,r)=\sin(qr)+\frac{1}{\Delta_{(0)}(q)}\bigg[V_1\sin(qa_2)G_{(0)}(q,r,a_1)
\end{equation}
\begin{displaymath}
+V_2\bigg(\sin(qa_2)+\frac{V_1}{q}\sin(qa_1)\sin[q(a_2-a_1)]\bigg)G_{(0)}(q,r,a_2)\bigg],
\end{displaymath}
\begin{eqnarray}
\label{nonr_ampl2}
f_{(0)}(q)=\frac{-\sin(qa_2)}{q^2\Delta_{(0)}(q)}\Bigg[\sum_{s=1}^{2}V_s\sin(q
a_s)+\frac{V_1V_2}{q}\sin(qa_1)\sin[q(a_2-a_1)])\Bigg],
\end{eqnarray}
where
\begin{eqnarray}
\label{det}\Delta_{(0)}(q)=1+\frac{1}{q}\sum_{s=1}^{2}V_s\exp(i q
a_s)\sin(q a_s)\\
+\frac{V_1V_2}{q^2}\exp(i q a_2)\sin(q a_1)\sin[q(a_2-a_1)].\nonumber
\end{eqnarray}
The non-relativistic limit of the quantization conditions
(\ref{qncn2_j}) takes the form
\begin{equation}
\label{nonr_qncn2}
\Delta_{(0)}(i\kappa)=1+\frac{1}{2\kappa}\sum_{s=1}^{2}V_s[1-\exp(-2\kappa
a_s)]
\end{equation}
\begin{displaymath}
+\frac{V_1V_2}{\kappa^2}\exp(-\kappa a_2)\sinh(\kappa
a_1)\sinh(\kappa(a_2-a_1))=0.
\end{displaymath}
All these expressions (\ref{nonr_wf2})-(\ref{nonr_qncn2}) coincide
with the expressions obtained by solving the Schr\"{o}dinger
equation with the superposition of two delta-shell potentials.

\section{Conclusion}
In this paper we solved exactly relativistic two-particle integral
equations describing the scattering and bound
$s$-states in cases of the delta-shell potential and the
superposition of two delta-shell potentials. These solutions yield
the expressions for the partial scattering amplitudes, partial
cross sections, partial $S$-matrices and phase shifts. The cross section behaviour demonstrates the resonant character of the scattering process for the potentials considered. Properties
of results obtained for the scattering and bound states are
investigated analytically and numerically. Namely: unitarity condition of the scattering amplitude is proved and the connection between the unitarity of scattering amplitudes and the imaginary part of the Green functions was found, the vanishing conditions of the
scattering amplitudes at some finite quantities of the rapidity is observed, the analysis of the conditions for bound state existence is performed. The non-relativistic limits of all obtained results are found.

The properties of solutions studied in this paper, and the physical values which were obtained based on these solutions have a general character for all four equations. In the future, we plan to consider a numerical solutions of  the two-particle integral equations for potentials which have more complicated form in the relativistic configurational representation.

\end{document}